\documentclass[conference]{IEEEtran}
\IEEEoverridecommandlockouts
\usepackage{amsmath,amssymb,amsfonts}
\usepackage{algorithmic}
\usepackage{graphicx}
\usepackage{textcomp}
\usepackage[table,xcdraw]{xcolor}
\usepackage{balance}
\usepackage{soul}
\usepackage{booktabs}
\usepackage{multirow}
\usepackage{float}
\usepackage{url}
\usepackage{tcolorbox}
\usepackage{array}
\usepackage{tabularx} 
\usepackage[noadjust]{cite}
\usepackage[bookmarks=true,pagebackref=true,hidelinks]{hyperref}
\usepackage{xurl}
\usepackage[flushmargin,para]{footmisc}

\usepackage{adjustbox}
\usepackage{tabularx}
\usepackage{caption}

\usepackage{listings} 
\usepackage{xcolor}   

\lstdefinestyle{javastyle}{
  language=Java,
  basicstyle=\ttfamily\small,    
  keywordstyle=\bfseries\color{blue},   
  commentstyle=\itshape\color{green!50!black}, 
  stringstyle=\color{red},   
  frame=single,               
  numbers=none,              
  numberstyle=\tiny\color{gray}, 
  stepnumber=1,              
  tabsize=4,                 
  showstringspaces=false,    
  breaklines=true,           
  showspaces=false,
  captionpos=b,
  rulecolor=\color{black}
}

\usepackage{enumitem}
\setlist[itemize,enumerate]{noitemsep, topsep=0pt, leftmargin=1.5em}
\usepackage[utf8]{inputenc}
\usepackage{newunicodechar,graphicx}
\DeclareRobustCommand{\okina}{%
  \raisebox{\dimexpr\fontcharht\font`A-\height}{%
    \scalebox{0.8}{`}%
  }%
}
\newunicodechar{ʻ}{\okina}
\usepackage{background}
\backgroundsetup{
  position=current page.east,
  angle=-90,
  nodeanchor=east,
  vshift=-3mm,
  opacity=1,
  scale=3,
  contents=PREPRINT
}

\begin{document}

\title{Identifier Name Similarities: An Exploratory Study}


\author{\IEEEauthorblockN{Carol Wong, Mai Abe, Silvia De Benedictis, Marissa Halim, Anthony Peruma}
\IEEEauthorblockA{University of Hawai‘i at Mānoa, Honolulu, Hawai‘i, USA \\
carolw8@hawaii.edu, maiabe@hawaii.edu, silviadb@hawaii.edu, mhalim@hawaii.edu, peruma@hawaii.edu}
}

\maketitle

\begin{abstract}
Identifier names, which comprise a significant portion of the codebase, are the cornerstone of effective program comprehension. However, research has shown that poorly chosen names can significantly increase cognitive load and hinder collaboration. Even names that appear readable in isolation may lead to misunderstandings in contexts when they closely resemble other names in either structure or functionality. In this exploratory study, we present our preliminary findings on the occurrence of identifier name similarity in software projects through the development of a taxonomy that categorizes different forms of identifier name similarity. We envision our initial taxonomy providing researchers with a platform to analyze and evaluate the impact of identifier name similarity on code comprehension, maintainability, and collaboration among developers, while also allowing for further refinement and expansion of the taxonomy.
\end{abstract}


\section{Introduction}
\label{Section:intro}
Identifiers, such as variable names, function names, and class names, play a vital role in source code. It is estimated that identifiers account for approximately 70\% of the total content in the codebase \cite{deissenboeck2006concise}. Developers typically spend around 55\% of their time understanding code \cite{xia2017measuring}, and high-quality identifier names can improve comprehension by up to 19\% \cite{hofmeister2019shorter}. Therefore, it is crucial for names to be both unambiguous and correctly reflect their intended behavior. 

Poor quality identifier names not only hinder code readability and degrade overall code quality (\cite{butler2009relating}) but can also lead to misinterpretations and increased cognitive load for developers \cite{Fakhoury2018}.  Prior research has examined how both the structural and semantic characteristics of names affect program comprehension. This includes aspects such as their length, case, grammatical structure, the use of abbreviations, acronyms, and digits, as well as how names evolve throughout the project's lifecycle \cite{lawrie2006whats,schankin2018descriptive,butler2015survey,peruma2022understanding,binkley-emse13,NewmanCoRR2020,NewmanICSME2019,PERUMA2020110704}. Through these studies, the research community has produced guidelines, best practices, and anti-patterns related to identifier naming as well as tools and techniques for identifier name appraisal and recommendations \cite{deissenboeck2006concise,Newman2022,Li2020Renamings,Alsuhaibani2022,Peruma2021,2021-ICPC-Methods}.

Despite the existing research on identifier naming (\cite{herka2023identifier, al2022would, newman2017lexical, gresta2020contextual}), there is limited work on understanding how the similarity between identifier names within a project impacts both code readability and developer productivity. Identifiers having the same or similar names but differing in purpose can confuse developers, resulting in potential bugs, increased cognitive load, and loss of productivity as developers try to decipher the true intent behind the name. This issue is especially pronounced in large teams, where developers may unintentionally reference the wrong identifier due to a similar name, or when onboarding new developers who are unfamiliar with the codebase \cite{arnaoudova2016linguistic}. 

\begin{lstlisting}[style=javastyle, caption=Same identifier name used in different contexts., label={lst:intro_code01}]
final OutputStreamWriter writer = new OutputStreamWriter(dataBuffer.asOutputStream().charset);

final FastString writer = new FastStringWriter(200);
\end{lstlisting}

Consider the code snippet in Listing \ref{lst:intro_code01}. Within the same project are multiple instances of an identifier with the same name (i.e., \texttt{writer}), yet they are used for different purposes and in different contexts. While both identifiers are related to writing data, their specific usage and functionality are distinct. This overlap can lead to misunderstandings, especially among novice developers who are still acclimating to the codebase.

While code clones (\cite{Baxter98}) have been explored extensively in the field \cite{Shobha2021,Kaur2023,ZakeriNasrabadi2023}, less attention has been given to naming similarities among identifiers. Identifiers, such as variable names, method names, and class names, are often included in code similarity measurement techniques \cite{ZakeriNasrabadi2023}. 

However, the specific study of similarities in naming among identifiers has not been thoroughly examined. Investigating identifier name similarities can significantly improve the quality of identifiers by revealing patterns of redundancy, inconsistency, ambiguity, and overall poor naming quality. The research will advance the body of knowledge in naming practices and how developers create and interpret names.

\subsection{Goal \& Research Questions}  
This paper presents an exploratory study aimed at improving our understanding of naming practices for identifiers by focusing on similar identifier names within a project. Our goal is to \textit{systematically identify and classify patterns of similarity in naming. We aim to construct an initial taxonomy that captures the diverse ways in which names resemble each other across real-world codebases}. We envision that this early-stage research will serve as a foundation for more comprehensive studies into the factors influencing the occurrences of such names and their impact on code comprehension and quality. Hence, we address the following research questions (RQ): 

\textbf{RQ 1: What types of similarities exist among identifier names, and how can they be effectively categorized?} Through this RQ, we aim to establish an initial taxonomy for identifier name similarities. By constructing a well-defined taxonomy, we can better understand the patterns of these similarities. Additionally, this classification will provide the research community with a platform for extending the taxonomy and the development of detection and analysis tools to aid in code quality assessment and refactoring. 

\textbf{RQ 2: How frequently do various categories of identifier name similarities occur in real-world software projects?} To complement the development of the taxonomy, this RQ aims to empirically examine the prevalence of each defined category of the taxonomy. By quantifying how often each category appears across open-source Java projects, we seek to understand which patterns are more common in practice. This analysis provides insights into which types of identifier similarities are most pervasive, potentially indicating areas of concern.


\subsection{Contributions}  
This study contributes to the field of identifier naming by:  
\begin{itemize}  
    \item Introducing a new taxonomy for classifying similarities in identifier names.
    \item Analyzing and quantifying patterns of naming similarity across open-source Java projects.  
\end{itemize}  

\section{Methodology}
\label{Section:method}
This section outlines the key activities in our methodology.

\subsection{Project Selection}
We used the SEART Data Hub web application \cite{Dabic:msr2021data} to identify open-source Java projects hosted on GitHub for our analysis. The tool enabled us to search for projects that met several criteria: a minimum of 10 stars, at least 500 commits, a minimum of 10 contributors, an age of at least two years, not being a fork, and a latest commit date between December 2024 and February 2025. These criteria are based on previous research (\cite{Dabic:msr2021data,eliseeva2024commit}) aimed at selecting popular projects with a high level of activity and developer involvement, which will strengthen the insights gained from our analysis.

As this is an exploratory study, we selected five random projects from the search results returned by the search tool. Though this sample size is limited, prior research has found that a saturation of themes can be found within a small number of cases \cite{hennink2017code}. The projects we selected to manually analyze are 
Dromara Sureness \cite{dromarasureness}, Spring Petclinic \cite{petclinic}, Netflix Metacat \cite{metacat}, Thymeleaf \cite{thymeleaf}, and  Apache Nutch \cite{apachenutch}.

\subsection{Identifier Name Extraction}
To facilitate analysis, we compiled an inventory of identifiers for each project. We used JavaParser to parse the (non-unit test) source files by traversing the Abstract Syntax Tree to ensure accuracy and consistency in the data collection.  The extracted identifiers include variables, parameters, classes, enums, and methods, along with their metadata: project name, file path, identifier name, data type (if applicable), the class and method in which the identifier was found, and the line number within its respective file. We identified 1,697 identifiers in Dromara Sureness, 494 in Spring Petclinic, 15,688 in Metacat, 21,876 in Thymeleaf, and 7,590 in Apache Nutch.

\subsection{Analysis Process}
The analysis began with each reviewer independently examining a specific project. Although identifiers were extracted using automation, the analysis of their semantic similarity was performed manually. This approach was necessary as existing automated techniques are limited in their ability to recognize semantic and contextual relationships within identifier names (\cite{wainakh2021idbench,al2022namesake}). In this study, the term ``semantic meaning'' refers to the real-world concept or abstraction that an identifier represents. For instance, the semantic meaning of the identifier ``account'' refers to a user's account within an application.

The reviewers analyzed sets of potentially similar identifiers by inspecting the surrounding code, verifying whether pairs of identifiers had consistent semantically similar meanings. For each pair, the reviewers documented identifier names, a proposed similarity category label, and contextual information about how each identifier was used within the code. After the initial review of a project, the projects were assigned to a second reviewer for an independent evaluation. This process helped ensure consistency and reduce individual bias.


Once all individual reviews were completed, the team compared their findings and refined the categories. Through iterative discussions, the reviewers resolved any discrepancies until they reached review saturation. At review saturation, additional reviews no longer provided new insights, and no significant changes were made to the taxonomy. A final taxonomy consisting of seven distinct categories was established. 

Using the finalized taxonomy, the team calculated the number of occurrences for each category for all reviewed projects. The distribution of these categories was analyzed to identify trends in identifier usage, allowing a comparative analysis of how similar identifiers are literally or semantically duplicated across different projects.

In each project, the number of identifiers analyzed exceeded the sample size necessary for statistical significance at a 95\% confidence level with a 5\% margin of error. This ensures that our findings are representative of the overall identifier population in each project. The number of identifiers analyzed was as follows: Spring Petclinic - 295, Metacat - 837, Sureness - 1,167, ThymeLeaf - 1,206, and Apache Nutch - 724.



\section{Results}
\label{Section:results}

\subsection*{\textbf{RQ 1: What types of similarities exist among identifier names, and how can they be effectively categorized?}}
By systematically analyzing five open-source Java projects, we identified seven distinct categories of identifier name similarities. 
Below, we provide a detailed explanation of each category, including its definition, implications, and code examples. Additionally, where relevant, some categories are further divided into subcategories to capture more nuanced patterns observed during our analysis.


\subsubsection{\textbf{Standardized Repetitive Names}}
Identifiers that intentionally reuse the same name across different scopes (methods, classes, files) to represent semantically equivalent entities or fulfill identical roles. This pattern reflects deliberate standardization and promotes consistency across the codebase. In our analysis, this pattern most commonly appeared in the form of standardized parameter names reused across methods for similar entities, constants and configuration keys consistently referenced across files, and local variables repeated within replicated code segments.

In Listing \ref{lst:StandardizedRepetitiveNamesExample}, the identifier "viewClass" is used in two separate documents to support configurable view resolution. However, reusing the same name for different rendering models can lead to confusion about the expected behavior. This ambiguity may result in developers misinterpreting the rendering context, introducing subtle bugs. 

\begin{lstlisting}[style=javastyle, caption=Standardized Repetitive Names Example,
label={lst:StandardizedRepetitiveNamesExample}]
private Class<? extends AbstractThymeleafView> viewClass = ThymeleafView.class;   

private Class<? extends ThymeleafReactiveView> viewClass = ThymeleafReactiveView.class;
\end{lstlisting}

\subsubsection{\textbf{Inconsistent Semantic Names}}
Different identifiers that perform the same semantic function or represent the same type of entity, creating inconsistency in naming conventions. This inconsistency can diminish code readability and increase cognitive load for developers. Such inconsistencies may not always be easily identified through automated tools and often necessitate manual review to ascertain the shared intent behind the 'identifiers. In Listing \ref{lst:InconsistentSemanticName}, the identifiers ``db'' and ``conn'' are utilized interchangeably to denote a database connection, but the name ``db'' fails to convey that it specifically represents a connection. This lack of clarity can obstruct developers’ understanding and disrupt the cohesiveness of the naming conventions throughout the codebase. 

\begin{lstlisting}[style=javastyle, caption=Inconsistent Semantic Names Example, label={lst:InconsistentSemanticName}]
public static synchronized boolean shrinkConnectionPoolSize() {
    ...
    DBConnection conn = null;
    ...
}

private static synchronized int getFreeDBConnectionNumber() {
    ...
    DBConnection db = null;
    ...
}
\end{lstlisting}

\subsubsection{\textbf{Colliding Names}}
Identifiers sharing lexically identical or similar names while representing completely different concepts or serving unrelated purposes. Although such identifiers may be lexically identical or closely related, their divergent semantic roles can introduce ambiguity and cause confusion.
This pattern was often observed in cases when developers favor brevity or reuse familiar terms across unrelated parts of the codebase. In Listing \ref{lst:NameCollisions}, a loop variable ``child'' appears in different contexts with unrelated semantic meanings, complicating developers' understanding from the names alone.
\vfill\null
\begin{lstlisting}[style=javastyle, caption=Colliding Names Example,label={lst:NameCollisions}]
if (classes != null) {
    for (File child : classes) {
        ...
    }
}
...
if (dirs != null) {
    for (File child : dirs) {
        ...
    }
}
\end{lstlisting}

\subsubsection{\textbf{Type-Based Variants}}
These are identifiers that are either lexically identical or similar, used within the same context but differing in their data types. These identifiers often represent the same conceptual entity in various forms and are utilized for transformations, parsing, or layered abstraction. While they may appear functionally similar, differences in data types can lead to confusion and increase the risk of misunderstandings during code maintenance.

\noindent\textbf{\textit{Polymorphic Names:}} This occurs when one identifier has a data type that is a subclass of another. This naming style can create ambiguity, especially if the names are similar or the type distinctions are unclear. For example, Listing \ref{lst:Type-VariantPolymorphic} has two parameters named "request": one for the type ``HttpServletRequest'' and its parent, ``ServletRequest''.

\begin{lstlisting}[style=javastyle, caption=Polymorphic Names Example, label={lst:Type-VariantPolymorphic}]
private static void addUserToSession(final HttpServletRequest request) {
    ...
}
...
private void doFilter(final ServletRequest request, final ServletResponse response, final FilterChain chain) throws IOException, ServletException {
    ...
}
\end{lstlisting}

\noindent\textbf{\textit{Cardinality Names:}} Identifiers that share similar names can sometimes lead to confusion when they differ in their cardinality, i.e., representing either a single value or a collection of values. In Listing \ref{lst:Type-VariantCardinality}, the variable ``agentName'', which holds a single string, and ``agentNames'', which contains a set of strings. A developer might misread the variable ``agentNames'' as ``agentName'' and overlook the critical difference indicated by the additional character `s'. 
\begin{lstlisting}[style=javastyle, caption=Cardinality Names Example, label={lst:Type-VariantCardinality}]
String agentName = conf.get("http.agent.name");
...
agentNames = new LinkedHashSet<>();
if (!agentName.equals("*")) {
  ...
  agentNames.add(agentName.toLowerCase());
}
\end{lstlisting}

\subsubsection{\textbf{Derivational Variants}}
These refer to identifiers that are deliberately named to reflect their derivation from a base or root identifier, particularly when behaviors change. These identifiers often appear in the same class or method. This pattern helps distinguish different stages of data transformation, temporary storage, or type specificity while maintaining a coherent naming relationship with the original identifier. However, overusing distinguishing names can increase cognitive load for developers, who must keep track of multiple variations of a concept. Additionally, as code evolves, names that were once clear can become ambiguous. In such cases, developers may need to refactor the code to rename identifiers, such as when a name includes a type that changes over time.

\noindent\textbf{\textit{Transformation Names:}} Identifiers that undergo transformations, such as parsing, trimming, filtering, or formatting. Typically, the naming convention for these identifiers includes a prefix or suffix that indicates the nature of the transformation. For example, in Listing \ref{lst:Semantic-Differences-Processed}, the variable ``input'' is transformed into ``scannedInput'', a lowercase version of the original string.
\begin{lstlisting}[style=javastyle, caption=Transformation Names Example, label={lst:Semantic-Differences-Processed}]
final String input = state.get(nodeIndex).getInput();
...
String scannedInput = input.toLowerCase();
\end{lstlisting}

\noindent\textbf{\textit{Type-Descriptive Names:}} Identifiers that are named similarly to their base names but include additional information about their data types. In this context, the descriptive identifier provides a more specific version of the base name. In Listing \ref{lst:Semantic-Differences-Type}, both ``target'' and ``targetObject'' are of type ``Object'', with ``targetObject'' explicitly indicating its data type.
\begin{lstlisting}[style=javastyle, caption=Type-Descriptive Names Example, label={lst:Semantic-Differences-Type}]
public boolean canRead(final EvaluationContext context, final Object target, final String name) throws AccessException {
    ...
}

public boolean canRead(final EvaluationContext context, final Object targetObject, final String name) throws AccessException {
    ...
}
\end{lstlisting}

\noindent\textbf{\textit{Temporary Names:}} Identifiers for short-lived variables that store temporary values. These variables are typically created to assist with validation, transformation, or processing of data, and are often discarded after their immediate use within a local scope. In Listing \ref{lst:Semantic-Differences-Temp}, the variable ``rolesTmp'' is a temporary variable that stores HTTP session attributes before being assigned to ``roles''. 

\begin{lstlisting}[style=javastyle, caption=Temporary Names Example, label={lst:Semantic-Differences-Temp}]
Object rolesTmp = httpSession.getAttribute(SurenessConstant.ROLES);
List<String> roles = rolesTmp = null ? null : (List<String>) rolesTmp;
\end{lstlisting}

\subsubsection{\textbf{Numerically Distinguished Variants}}
These refer to identifiers that include numbers to differentiate between distinct instances or roles of the same underlying concept. This technique helps prevent naming conflicts but may also indicate opportunities for better abstraction through the use of collections or arrays. 

\noindent\textbf{\textit{Sequential Numeric Names:}} Identifiers that append numeric-suffixes to differentiate multiple objects of the same type. For example, in Listing \ref{lst:DigitDistinguishedVariants}, when managing multiple ``Customer'' instances, the developer names each instance as ``cust1'', ``cust2'', and ``cust3'', each representing a unique customer.  

\begin{lstlisting}[style=javastyle, caption=Sequential Numeric Names Example, label={lst:DigitDistinguishedVariants}]
final Customer cust1 = new Customer();
...
final Customer cust2 = new Customer();
...
final Customer cust3 = new Customer();
\end{lstlisting}

\noindent\textbf{\textit{Value-Encoded Names:}} Identifiers that encode specific values directly in their name, often for clarity or to enforce specific mappings. In Listing \ref{lst:ValueEncodedVariant}, the identifiers ``COUNT\_2`` and ``COUNT\_3`` directly indicate their associated value. This is not inherently sequential, as the absence of ``COUNT\_1`` indicate that the numbers were chosen for their semantic value. 

\begin{lstlisting}[style=javastyle, caption=Value-Encoded Names Example, label={lst:ValueEncodedVariant}]
private static final int COUNT_2 = 2;

private static final int COUNT_3 = 3;
\end{lstlisting}

\subsubsection{\textbf{Concise Variants}}
These are identifiers that utilize shortened or compact naming forms to convey their meaning. These names often prioritize brevity over descriptiveness. 

\noindent\textbf{\textit{Abbreviated Names:}} These are identifiers that represent shortened forms of longer, more descriptive terms. It is important to use abbreviated names cautiously to maintain readability, especially in larger or long-lived codebases. In Listing \ref{lst:ConciseAbbreviated}, the variables ``logger'' and ``log'' both record messages in the console, but their similar naming may cause confusion regarding any functional differences between them.
\begin{lstlisting}[style=javastyle, caption=Abbreviated Names Example, label={lst:ConciseAbbreviated}]
public static final Logger log = LoggerFactory.getLogger(HandlerManager.class);

public static final Logger logger = LoggerFactory.getLogger(HandlerManager.class);
\end{lstlisting}

\noindent\textbf{\textit{Acronym Names:}} These are identifiers formed by the initial letters of compound terms or multi-word phrases, typically written in uppercase or camel case. In Listing \ref{lst:ConciseAcronym},the variable named ``sb'' is an acronym for StringBuilder.
\vfill\null
\begin{lstlisting}[style=javastyle, caption=Acronym Names, label={lst:ConciseAcronym}]
private String calcDigest(String first, String ... args) {
    StringBuilder stringBuilder = new StringBuilder(first);
    ...
}
private static String bytesToHexString(byte[] bytes) {
    StringBuilder sb = new StringBuilder();
}
\end{lstlisting}

\noindent\textbf{\textit{Single-Character Names:}} Single-character identifiers are often used in tightly scoped contexts, such as mathematical operations, iterations, or lambda functions. However, in some cases, these short names are applied to more complex objects, leading to potential readability issues. In Listing \ref{lst:ConciseSingle}, a byte array is declared with the name ``b'', while a StringBuffer variable is also named ``b''. In this scenario, the use of brief name makes it less clear what each variable represents.

\begin{lstlisting}[style=javastyle, caption=Single-Character Names Example, label={lst:ConciseSingle}]
byte[] b = new byte[1024];

StringBuffer b = new StringBuffer();
\end{lstlisting}

\subsection*{\textbf{RQ 2: How frequently do various categories of identifier name similarities occur in real-world software projects?}}

\begin{table*}[t]
\centering
\caption{Identifier Similarities Count.}
\label{RQ2_Table}
\begin{adjustbox}{max width=\textwidth}
\begin{tabular}{p{2cm} p{2cm} p{4cm} p{4cm} p{4cm}} 
\textbf{Project} & \textbf{Identifier Similarities Count} & \textbf{Top Category} & \textbf{2nd Category} & \textbf{3rd Category} \\
\hline
Spring Petclinic & 267 (90.5\%) & Standardized Repetitive Names (56.18\%) & Concise Variants - Abbreviated (16.86\%) & Colliding Names (16.10\%) \\
Metacat & 517 (61.77\%) & Concise Variants - Abbreviated (70.41\%) & Derivational Variants - Type-Descriptive Names (13.73\%) & Standardized Repetitive Names (7.93\%) \\
Dromara Sureness & 129 (11.05\%) & Standardized Repetitive names (40.31\%) & Type-Based Variants - Cardinality Names (13.18\%) & Numerically Distinguished Variants (10.08\%) \\
Thymeleaf & 291 (24.13\%) & Standardized Repetitive Names (79.38\%) & Name Collisions (7.22\%) & Type-Based Variants - Cardinality Names (3.43\%) \\
Apache Nutch & 93 (12.85\%) & Concise Variants - Single-Character (47.31\%) & Numerically Distinguished Variants (27.96\%) & Standardized Repetitive Names (18.28\%) \\
\end{tabular}
\end{adjustbox}
\end{table*}

Identifier name variant counts and the three most common variant categories for each project are shown in Table \ref{RQ2_Table}.

Spring Petclinic was found to have the highest percentage of identifiers that were classified as variants with 267 (90.5\%) variants found while Apache Nutch had the lowest percentage of identifiers variants with 93 (12.85\%) instances. A higher percentage of identifier variance points to a more uniform, possibly template-driven naming structure across the codebase. 

Standardized Repetitive Names is among the top three most common categories across all projects, suggesting that most codebases deliberately attempt to maintain naming consistency. Spring Petclinic's high variant percentage suggests that Spring Petclinic has a smaller, more homogenous project structure or relies more on boilerplate code. This contrasts with Dromara Sureness and Apache Nutch, where their lower percentage of identifier name similarities may indicate diverse functionality in the project or domain-specificity.

Projects with high variant percentages can have improved readability and easier onboarding due to its uniformity. However, they may have less functional diversity with smaller-scope projects. In contrast, projects with low variant percentages may have a broader functional scope, but may be harder to consistently enforce, leading to reduced clarity in the code. Identifier variants can reflect the functional complexity and scale of a codebase, guiding decisions on refactoring or the onboarding strategy.

\section{Discussion}
\label{Section:discussion}

Our exploratory study examined patterns of identifier name similarities across five widely used Java projects and introduced a taxonomy comprising seven categories, some of which have further subcategories. Our analysis revealed that many identifier similarities were driven by variations in data types and by differences in how identifiers were used within their respective contexts. Aspects of an identifier's semantic features, such as its role, data structure, and scope, were critical in understanding the intent and function of the naming choices. Identifier pairs declared close to one another were more likely to be semantically related, even when named differently. This suggests that automated detection tools must consider contextual information beyond simple string matching to provide meaningful similarity assessments.

During manual inspection, a considerable number of abbreviated or shorthand identifiers were encountered, particularly within utility classes or method-scoped logic. This suggests that developers often prioritize brevity over clarity when naming identifiers, especially in localized scopes \cite{hofmeister2019shorter}. Although these abbreviations, such as ``u'' for url or ``c'' for Comparator or Character, are syntactically concise, they often obscure the underlying intent, especially when reused across files or contexts. This made semantic tracing more difficult, highlighting a potential readability and maintainability risk in using excessively concise names. 

Another unexpected finding was the discovery of inconsistently named identifiers that serve the same purpose, even in well-established and widely used open-source projects. For example, in Dromara Sureness, ``clazz`` and ``clz`` are both used as an iterator variable to loop through different object types. Similarly, in Thymeleaf, the identifiers ``target`` and ``element`` were both used to refer to an iterable target object, and were named differently despite having no structural differences.


The patterns identified in this study have clear implications for software maintainability. However, an important finding is that not all identifier similarities represent problems.
Certain categories, such as Colliding Names, Inconsistent Semantic Names, and Concise Variants, pose a greater risk to readability, especially in large codebases. These patterns increase the cognitive load on developers, who need to understand the naming ambiguities without explicit documentation or naming conventions. Moreover, when identifiers are semantically different but share the same or similar names, they can introduce bugs or reduce the effectiveness of automatic refactoring and static analysis tools. 
Conversely, patterns such as Derivational Variants and Type-Based Variants are often named intentionally similar to provide clarity, especially when used to indicate stages of data transformation or temporary roles.

\vspace{1.5mm}
\noindent{\textbf{\textit{Implications for Researchers:}}} Our taxonomy provides a platform for investigating identifier naming practices, specifically similar names in the codebase. This work opens up several research opportunities, including large-scale empirical studies on more diverse and representative projects to examine the occurrence of similarity categories, the cognitive impact of different similarity types, investigations into how such names relate to code quality metrics, and refining existing code readability models to account for similar identifier names. Moreover, the taxonomy can serve as a foundation for developing naming guidelines and refactoring recommendations.

\vspace{1.5mm}
\noindent{\textbf{\textit{Implications for Developers:}}} Our findings suggest that teams should establish explicit naming conventions that account for the different types of similarities we identified. By formalizing these naming conventions, development teams can enhance code readability and reduce cognitive load when navigating large codebases. These insights can improve code reviews, automated tools, and onboarding documentation, leading to improvements in software quality and team productivity. 

\vspace{1.5mm}
\noindent{\textbf{\textit{Implications for Tool/IDE Vendors:}}} Using our findings, tool and IDE vendors can develop automated tools that can identify and classify different types of naming similarities. Furthermore, integrating such tools into IDEs can provide developers with real-time feedback on their naming choices or into code review systems to flag potentially problematic names. Advanced features could include suggesting alternative identifier names tailored to the surrounding code context or visualizing naming similarity clusters across a project.

\vspace{1.5mm}
\noindent{\textbf{\textit{Implications for Software Engineering Education:}}} Traditional programming education often focuses on basic naming conventions, yet challenges of naming similarities in larger codebases are not typically addressed. The taxonomy can help students recognize and avoid problematic naming patterns. Incorporating these concepts into programming courses can provide future developers with strategies for writing more readable and maintainable code in complex environments.

\section{Threats To Validity}
\label{Section:threats}
The findings from this study are limited to our analysis of only five open-source Java projects, which threatens the generalizability of our findings. However, as an exploratory study, the findings serve as a starting point for further research in this area. In addition, the projects that were chosen were larger codebases that had many developers contributing to them. Future studies could expand the sample size to include a broader range of projects across different sizes, domains, paradigms, and programming languages. 

Additionally, our manual analysis of the code may introduce subjectivity, as the reviewers need to interpret the code's context and the developer's intent when classifying similar identifiers. This subjectivity could impact the reliability of our results due to potential inconsistencies in categorization. To mitigate this threat, we employed multiple reviewers for each project, with reviewers cross-validating results by swapping projects and resolving any disagreements through discussion.

Finally, our taxonomy has not been validated by practitioners, which threatens its validity. Without input from practitioners, there is a risk that the categories may not accurately reflect real-world practices or challenges that developers face.   

Despite these limitations, this exploratory study offers valuable insights and establishes a foundation for future research. Moreover, the taxonomy should be viewed as a starting point that requires further refinement and validation across different domains, programming languages, and development contexts.




\section{Conclusion \& Future Work}
\label{Section:conclusion}
Motivated by the important role that identifiers play in program comprehension and maintainability, this study explores the prevalence and categorization of similar identifier names in a sample of open-source Java projects. We have identified a diverse range of naming patterns and developed a novel taxonomy that includes multiple categories and several subcategories. Our work advances the existing knowledge in the field of identifier naming and serves as a foundational step towards a better understanding of the presence of similar identifier names within a project's codebase.

Our future work will refine this taxonomy by incorporating additional projects across different programming languages. This will allow us to create a more effective and generalizable taxonomy, as well as identify any language-specific or domain-specific naming trends. Additionally, we plan to engage with practitioners to validate our taxonomy and gain insights into the practical implications of similar identifier names, such as their effects on code readability and maintenance efforts.

\vspace{2mm}
\noindent \textit{Artifacts related to this study are available at \cite{website}.}


\bibliographystyle{ieeetr}
\bibliography{main}

\end{document}